\documentclass[prl,twocolumn,showpacs,superscriptaddress,
preprintnumbers,amsmath,amssymb,tightenlines]{revtex4}
\usepackage{graphicx}
\usepackage{bm}% bold math
\usepackage{graphicx}% Include figure files
\newcommand{\bea}{\begin{eqnarray}}
\newcommand{\eea}{\end{eqnarray}}

\newcommand{\ba}{\begin{array}}
\newcommand{\ea}{\end{array}}
\newcommand{\bc}{\begin{center}}
\newcommand{\ec}{\end{center}}

\newcommand{\bml}{\begin{mathletters}}
\newcommand{\eml}{\end{mathletters}}
\newcommand{\commentout}[1]{{}}

\newcommand{\half}{\hbox{$1\over2$}}

\newcommand{\eq}[1]{(\ref{#1})}

\begin{document}

\title{Optical detection of fractional particle number in an atomic
Fermi-Dirac gas}
\author{Juha Javanainen}
\affiliation{Department of Physics, University of Connecticut, Storrs, CT
06269, USA}
\email{jj@phys.uconn.edu}
\author{Janne Ruostekoski}
\affiliation{Department of Physical Sciences, University of
Hertfordshire, Hatfield, Herts, AL10 9AB, UK}
\email{j.ruostekoski@herts.ac.uk}
\date{\today }
%\maketitle
\begin{abstract} We study theoretically a Fermi-Dirac atomic gas in a
one-dimensional optical lattice coupled to a coherent electromagnetic
field with a topologically nontrivial soliton phase profile. We argue that
the resulting fractional eigenvalues of the particle number operator can
be detected via light scattering. This could be a truly quantum
mechanical measurement of the particle number fractionalization in a
dilute atomic gas.
\end{abstract}

 \maketitle

It has been known for a while that, in the presence of a
topologically nontrivial bosonic background field, fermionic
particles may carry a fractional part of an elementary quantum
number~\cite{JAC76,NIE86}. In the condensed matter regime this
phenomenon was introduced~\cite{SU79} to describe conjugated
polymers. The existence of fractionally charged excitations in the
polymers is typically demonstrated indirectly by detecting the
reversed spin-charge relation~\cite{HEE88}. The fractional quantum
Hall effect (FQHE) can also be explained by invoking
quasiparticles, each with a fraction of an electron's
charge~\cite{laughlin}. The fluctuations of the tunneling current
in low-temperature FQHE regime have been measured~\cite{DEP97}.
Interpreting the current shot noise according to the
Johnson-Nyquist formula duly suggests that the current is carried
by the fractional Laughlin quasiparticles. Analogous experiments
have determined the fractional expectation value of the charge in
FQHE in the Coulomb blockade regime~\cite{GOL95}.

We have earlier proposed a system of Fermi-Dirac (FD) atoms in an
optical lattice that should display fractional atom
numbers~\cite{RUO02}. In the present Letter we argue that
usual optical methods such as phase contrast imaging and
measurements of the intensity of light scattered by the atoms
extract information about the fractional fermion number. The
technical challenges are severe, but in principle both the
expectation value and the fluctuations of the atom number are
accessible to experiments.

Briefly, we consider a FD atom \cite{OLS98} with a $\Lambda$
scheme for two active states (say, Zeeman states in different
hyperfine levels) that can be coupled by one or two-photon
electromagnetic (em) transitions~\cite{RUO02}. The atoms reside in
a 1D optical lattice \cite{lattices} that holds the two states at
alternating sites $\lambda/4$ apart, where $\lambda$ denotes the
wavelength of lattice light. By making use of the em transitions,
we assume, it is possible to make the atoms hop between the
adjacent sites so that they at the same time change their internal
state. The lattice Hamiltonian is
\begin{equation}
   {H\over\hbar} = \sum_k \left[\delta_k c^\dagger_kc_k
+\kappa_k(c^\dagger_{k+1}c_k + c^\dagger_k c_{k+1})\right]\,.
\label{HAM}
\end{equation}
   Here $c_k$ is the annihilation operator for the fermionic
atoms at site
$k$, $\hbar \delta_k$ is the energy of the atoms at
$k$, and $\kappa_k$ are the are hopping matrix elements tailored to suit
our purposes. The literature is replete with studies of similar
systems~\cite{BEL83,KIV,GW}, but our treatment is unusual in that we never
resort to the continuum limit.

It is easy to see that the matrix
$U_{kp}$ whose columns are the orthonormal eigenvectors of the eigenvalue
problem
\begin{equation} (\delta_k -\omega)a_k +  \kappa_k a_{k+1} +
\kappa_{k-1} a_{k-1} = 0
\label{EVP}
\end{equation} may be employed to diagonalize the Hamiltonian~(\ref{HAM}).
In terms of the new fermion operators $\gamma_p\equiv \sum_k
U_{kp} c_k$, we have $H/\hbar=\sum_p \omega_p \gamma^\dagger_p
\gamma_p$ . Without a loss of generality, the couplings $\kappa_k$
are assumed real, and so we take $U$ real and orthogonal;
$U_{kp}^{-1}=U_{pk}$. In this paper all calculations are done
directly numerically.

Even though the optical lattice may be part of a larger trap and the
wider trap could generate interesting physics in its own right, we
simplify by putting $\delta_k=0$. The fractional charge arises from
certain types of defects in the couplings $\kappa_k$. We illustrate by
assuming a {\it dimerized} lattice generated by the coupling matrix
element that alternates from site to site between two values $a+\mu$ and
$a-\mu$, except that at the center of the lattice there is a defect such
that the same coupling matrix element appears twice. We take
the number of lattice sites to be $N_s\equiv 2N_h+1\equiv 4n+1$, where
$n$ is an integer itself, and number the sites with integers ranging from
$-N_h$ to $N_h$.  For illustration, pick $n=2$, use an {\rm x} to denote
a lattice site, and $\pm$ the couplings
$a\pm\mu$, then our lattice with the couplings reads
\begin{equation} {\rm x}+{\rm x}-{\rm x}+{\rm x}-{\rm x}-{\rm x}+{\rm
x}-{\rm x}+{\rm
x}\,
\end{equation}

It is then easy to see from the structure of Eq.~\eq{EVP} that if
$\omega$ is an eigenvalue, then so is $-\omega$; and the eigenvectors
transform into one another by inverting the sign of every second
component. We will label the eigenvectors as
$-N_h\ldots N_h$ in ascending order of frequency, and assign the
labels $\pm p$ to such $\pm$ pair of states. Correspondingly, the
transformation matrix $U$ satisfies $|U_{kp}| = |U_{k,-p}|$.

But under our assumptions, the number of eigenvalues and eigenstates is
odd. The $\pm$ symmetry implies that an odd number of the eigenvalues
must equal zero. Except for special values of the couplings $a$ and $\mu$,
there is one zero eigenvalue. We call the corresponding eigenstate the
zero state. Provided $a$ and $\mu$ have the same sign and $|a|>|\mu|$,
all odd components in the zero state equal zero and the even components
are of the form
\begin{equation} x_k = x_0\left({-\displaystyle{a-\mu\over
a+\mu}}\right)^{|k|/2}\,.
\end{equation} The zero state becomes the narrower, the closer in
absolute value
$a$ and $\mu$ are.

The dimerized optical lattice resulting from the alternating pattern of
the hopping matrix elements causes the single-particle density of states
to acquire an energy gap, which  in the limit $N_s\rightarrow\infty$
equals $4\hbar |\mu|$ . The zero state is located at the center of the
gap. The resulting excitations at half the gap energy could be detected by
resonance spectroscopy. This provides indirect evidence of
fractionalization, as in the polymer systems~\cite{HEE88}. Because in our
scheme~\cite{RUO02} the gap is proportional to the amplitude of the em
field inducing the hopping, the size of the energy gap can be controlled
experimentally. The effective zero temperature limit, $|\mu|\gg
k_BT/\hbar$, might then be reached under a variety of experimental
conditions.

Suppose next that the system is at zero temperature, and contains
$N_f=N_h+1$ fermions. The exact eigenstates $p$ are then filled up to zero
state and empty at higher energies, with occupation numbers $n_p$ = 0 or
1.  The number operator for the fermions at site $k$ correspondingly reads
$c^\dagger_k c_k=\sum_{pq} U_{kp} U_{kq}\gamma^\dagger_p
\gamma_q$, so the expectation value of the fermion number at the site $k$
is
\bea \langle c^\dagger_k c_k \rangle  &=& \sum_{p=-N_h}^0 |U_{kp}|^2 =
\half\sum_{p=-N_h}^{N_h}|U_{kp}|^2 + \half|U_{k0}|^2\nonumber\\ &=& \half
+ \half|U_{k0}|^2\,.
\label{expnumber}
\eea The second equality is based on the symmetry $|U_{kp}| =
|U_{k,-p}|$, and the third on the orthogonality of the matrix $U$. By
virtue of the same orthogonality, localized with the zero state there is
a lump with $\half\sum_{k} |U_{k0}|^2=\half$ fermions on top of a uniform
background of half a fermion per site. This lump is
the celebrated half of a fermion. So far we only deal
with the {\it expectation\/} values of the atom numbers, but we will
demonstrate shortly that the {\it fluctuations\/} in atom number can be
small as well.

Fractionilization is a more robust phenomenon than our discussion may let on. Something akin to a localized zero state occurs as soon as the regular alternation of the couplings between adjacent states gets out of rhythm around a defect. In particular, the defect does not have to be confined to one lattice site, which might make the experiments easier. The half-fermion is localized, so it does not critically depend on the number or parity of sites, and not even on the exact number of the fermions. We will enumerate such variations of the theme elsewhere.

We next turn to the optical detection of the FD gas.  We assume
that far off-resonant light excites the atoms, whereupon the 1D
optical lattice may be considered optically thin.  We take the
sources residing at each site to be much smaller than the
wavelength of light. Then the (positive frequency part of the)
field operator for scattered light is~\cite{JAV95} $\hat{E}^+  =
C\sum_k \alpha_kc^\dagger_kc_k$, where $C$ is a constant
containing the overall intensity scale of the driving light.
Henceforth we scale  so that $C=1$. The factors $\alpha_k$ include
aspects such as intensity and phase profile of the driving light,
effects of the spin state at each site $k$ on the light-atom
coupling, and  propagation phases of light from the lattice site
to the point of observation.

In forward scattering and variations thereof such as phase contrast
imaging, the scattered and the incoming light interfere. The ultimate
measurement of the intensity in effect records the expectation value of
the electric field $E=\langle
\hat{E}^+\rangle$. The observable at the detector is
\begin{equation} E = \sum_k \alpha_k \langle c^\dagger_k c_k\rangle =
\sum_{k p}
\alpha_k U^2_{kp} n_p\,.
\end{equation} This is a linear combination of the expectation values of
the numbers of fermions at each lattice site with the coefficients
$\alpha_k$, which are to some extent under the control of the
experimenter. In the absence of interference with the incoming light, the
simplest observable is light intensity $I=\langle
\hat{E}^-\hat{E}^+ \rangle$. The detector then probes the quantity
characteristic of the FD statistic
\bea I &=& \sum_{kl} \alpha^*_k \alpha_l \langle c^\dagger_k c_k
c^\dagger_l c_l\rangle =|E|^2 + \Delta I\,; \label{FRS}\\ \Delta I &=&
\sum_{klpq} \alpha^*_k \alpha_l U_{kp}U_{kq}U_{lp}U_{lq}n_p(1-n_q)\,.
\label{FFLU}
\eea

We now construct a numerical example about the use of forward scattering to detect the fractional particle. We make use of the fact that the fermion species at the alternating lattice sites are likely to be different.
%We assume that the driving light is far blue-detuned in one species %and far red-detuned in the other, and that the two dipole matrix %elements are comparable. One may then find a laser tuning
We assume that a given driving light is far above resonance
with the fluorescing transition in one species and far below resonance in the other, and that the matrix elements for dipole transitions are
comparable. It is then possible to find a tuning of the laser
such that the intensity of scattered light is the same for both species. Moreover, the light scattered by the two species are out of phase by $\pi$, and out of phase with the incident light by
$\pi/2$. With the usual tricks of phase contrast imaging, the relative
phase of incident and scattered light is then adjusted so that that in
interference light from one species directly adds to the incident
light, and light from the other species subtracts.

The second element of the argument is a rudimentary model for an imaging
system with a finite aperture. Let us assume the geometry has been
arranged in such a way that all Fourier components of light in the plane
of the aperture up to the absolute value $K$ are passed, the rest
blocked. The effect on imaging from the object plane to image plane can
be analyzed by taking the Fourier transform of the object, filtering with
the multiplier $\theta(K-|{\bf k}_\perp|)$, and transforming back. The
filtered image of the lattice filtered is then proportional to
\begin{equation}
E({\bf r}) = \sum_k (-1)^k \langle c^\dagger_k
c_k\rangle\,{2 J_1(K |{\bf r}-{\bf r}_k|)\over
K|{\bf r}-{\bf r}_k|}\,.
\end{equation}

\begin{figure}
\includegraphics[width=7cm]{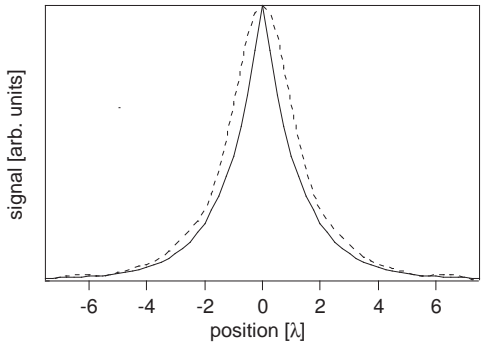}
\vspace{-10pt}
\caption{Optical image (dashed line) of a zero state carrying half
of a fermion (solid line) for a specific imaging system as
discussed in the text. The size of the soliton is set by the
choices $N_s=129$, $N_f=65$, $\mu=0.1\,a$.} \label{IMAGE}
\end{figure}

We choose the parameters $\mu = 0.1\,a$, and the numbers of sites
and fermions $N_s=129$ and $N_f=65$. We take the numerical
aperture $F=1$ for the imaging system, and the corresponding
maximum possible  cutoff wave number $K=2\pi/(\sqrt{5}\lambda)$.
In Fig.~\ref{IMAGE} we plot the optically imaged fermion lattice
along the line of the atoms (dashed line), and the number of
fermionic atoms in excess of the average occupation number \half\
for the even-numbered sites that carry the zero state (solid
line), as obtained from Eq.~(\ref{expnumber}). The curves are
normalized so that the maxima overlap. The imaging system picks up
a resolution rounded version the half-fermion hump.

In fact, phase contrast imaging has been used for nondestructive
monitoring of a Bose-Einstein condensate~\cite{AND96}, and the absorption
of a single trapped ion has been detected experimentally~\cite{ION80}.
While a lot of assumptions went into our specific example,
an experiment along these lines should be feasible with the technology
available today.

With illumination of the optical lattice by a focused
light beam and detection of scattered intensity in a direction of
constructive interference, it is in principle also possible to realize a
situation in which the weights approximately make a Gaussian distribution
around the zero state, $\alpha_k = e^{-(k/w)^2}$. In such a case the
observable $\hat N=\hat{E}^+$ is just a linear combination of the
occupation numbers of the lattice sites, the quantity $E$ is the
expectation value thereof, and
$\Delta I$ is nothing but the squared fluctuations of $\hat N$,
$\Delta I =(\Delta N)^2 = \langle{\hat N}^2\rangle - \langle {\hat
N}\rangle^2$.

\begin{figure}
\includegraphics[width=7.5cm]{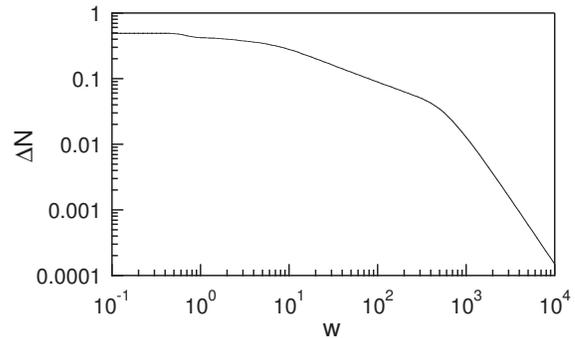}
\vspace{-10pt}
\caption{Fermion number fluctuations $\Delta N$ under a Gaussian
envelope of width $w$, for $N_s=1025$,  $N_f=65$,
$\mu=0.1\,a$. } \label{FLUCTS}
\end{figure}

To illustrate, we take a lattice with $N_s=1025$ sites, pick the
parameters $\mu=0.1\,a$, put in $N_f=513$ fermions so that the zero state
is the last filled stated, and find the rms fluctuation of the fermion
number $\Delta N$ as a function of the width of the weight function $w$.
The result is shown in Fig.~\ref{FLUCTS} on a log-log plot. The notch
around $w=1$ indicates that at this point the weight factors
$\alpha_k$ start to cover several lattice sites. Another break in the
curve is seen at about $w=10$, when the weight function covers the whole
zero state. Thereafter the fluctuations behave as $\Delta N\propto
w^{-1/2}$.  The fermion number $\hat N$ under the weight function becomes
more sharply defined as the region for averaging grows broader. Finally,
at $w$ of a few hundred, the weights $\alpha_k$ effectively cover the
entire lattice. The fluctuations then decrease even faster with
increasing $w$, as is appropriate for the fixed fermion number in the
lattice as a whole.
\begin{figure}
\includegraphics[width=7.0cm]{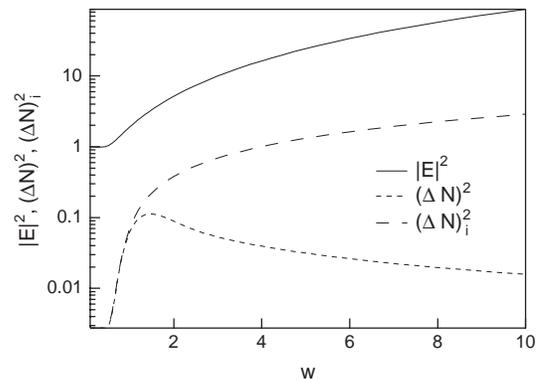}
\vspace{-10pt}
\caption{The intensity of light scattered from the optical lattice
if the fermion numbers did not fluctuate ($|E|^2$), and the
additional intensity due to fermion number fluctuations ($(\Delta
N)^2$), as a function of the size of the focus $w$ of the
driving light given in the lattice units of $\lambda/4$. We also
display the added intensity $(\Delta N)^2_i$ that would result if
fermion number fluctuations were uncorrelated between adjacent
lattice sites. The soliton parameters are $N_s=129$, $N_f=65$,
$\mu=0.9\,a$.} \label{INTENSITY}
\end{figure}

In the standard half-integer fermion number arguments one subtracts a
neutralizing background of precisely $\half$ charge per lattice site,
whereupon $\langle \hat{N}\rangle
\rightarrow\half$ and $\Delta N\rightarrow 0$ with an increasing width
$w$. The intermediate regime that occurs once the zero state is covered is the crux of the matter. Not only does the expectation value of fermion number equal $\half$, but fluctuations are also small. {After the subtraction}, the fermion number has the eigenvalue $\half$. From the quantum optics viewpoint, this is something of a conjuror's trick.
Correlated fluctuations in fermion number between adjacent sites create an impression of a sharp eigenvalue in a smoothly weighted sum of the occupation number operators for the lattice sites.

The scattered light carries signature of the fluctuations in the
scattered intensity.  We demonstrate by plotting in
Fig.~\ref{INTENSITY} separately the contribution $|E|^2$, as if
the fermion numbers were precisely fixed, and the fluctuation
terms $(\Delta N)^2$. We also show the quantity $(\Delta N)_i^2$
that remains from the fluctuation term if we only keep the contributions
with $k=l$ in Eq.~(\ref{FFLU}), as if the fermion number fluctuations at
adjacent sites were uncorrelated. These are given as functions of the
width of the focus $w$ of the laser beam. Here $N_s=129$, $N_f=65$, and
we choose $\mu=0.9\,a$ to make a sharply localized zero state.

Even with a very narrow focus of the laser, $w=4$ or about one
wavelength, the contribution from the fluctuations is two orders of
magnitude below the coherent intensity, whereas the fluctuations from
uncorrelated fermion numbers would make a contribution an order magnitude smaller than the coherent intensity. As our detection light was assumed to be far-off resonance, the photon number fluctuations are Poissonian.
Under otherwise ideal conditions, the detection of about a hundred
photons could therefore reveal the difference between correlated and
uncorrelated fermion numbers at adjacent site, whereas a quantitative
study of the actual correlated fermion numbers requires the detection of about 10,000 photons. Unfortunately, a large number of scattered photons means a large number of recoil kicks on the fermions. Currently available optical lattices likely cannot absorb the assault of hundreds of photon recoils without developing some form of a dynamics that complicates the phenomena we are analyzing. In the coherent detection of the fermion number, however, the photon recoils could be suppressed by tuning the energy gap to be much larger than the photon recoil energy.

It is instructive to note that, at the level we have discussed
(amplitude or intensity measurements), optical detection of the anomalously small fermion number fluctuations responsible for fractionilization has to be coherent and rely on
interference of light scattered from different lattice sites. If a too
broad angular average or other such cause wipes out the interferences
[$\alpha^*_k\alpha_l\rightarrow\delta_{kl}\alpha^*_k\alpha_k$], we are
back to adding fermion number fluctuations from different lattice sites as if they were independent.

Although we, of course, do not aim at a specific experimental
design, a few variations to potentially overcome the technical
limitations we have noted bear a mention. First, we have
implicitly assumed that the lattice light and the detection light
have the same wavelengths. By angling the beams used to make the
optical lattice, or possibly by using microlens
arrays~\cite{DUM02}, the optical lattice can be stretched. The
resolution limit imposed by the wavelength of the detection light
could be circumvented. Second, so far we have dealt with what in
essence is spontaneous Bragg scattering. Recently, induced Bragg
scattering has been introduced as a method to study the
condensates in detail~\cite{STA99}. How induced scattering works
in the case of a 1D lattice under inhomogeneous illumination is
not clear at the moment, but conceivably the Bragg pulses could
be made so short that the harmful effects of photon recoil do not
have time to build up during the measurement.

We have discussed optical detection of the half-fermion that can arise
from a topological defect in an optical lattice holding a FD gas. Even
though both the average fermion number and its fluctuations are in
principle amenable to optical measurements, experiments will evidently
have to await further development of technology. In the interim, the most
valuable outcome of the kind of an analysis we have presented would
probably be the insights it brings into the phenomenon of a fractional
fermion number and its prospective applications.

We acknowledge financial support from the EPSRC, NSF, and NASA.


\begin{references}

\bibitem{JAC76} R. Jackiw and C. Rebbi, Phys. Rev. D {\bf 13}, 3398
(1976).

\bibitem{NIE86} A. Niemi and G. Semenoff, Phys. Reports {\bf 135}, 99
(1986).

\bibitem{SU79} W.P. Su, J.R. Schrieffer, and A.J. Heeger, Phys. Rev.
Lett. {\bf 42}, 1698 (1979).

\bibitem{HEE88} A.J. Heeger {\it et al.}, Rev. Mod. Phys. {\bf 60}, 781
(1988).

\bibitem{laughlin} R. B. Laughlin, H. St\"ormer and D. Tsui, Rev. Mod.
Phys. {\bf 71}, 863 (1999).

\bibitem{DEP97} R. de-Picciotto {\it et al.}, Nature {\bf 389}, 162
(1997); L. Saminadayar {\it et al.}, Phys. Rev. Lett. {\bf 79},
2526 (1997).

\bibitem{GOL95} V.J. Goldman and B. Su, Science {\bf 267}, 1010 (1995).

\bibitem{RUO02} J. Ruostekoski, G. V. Dunne, and J. Javanainen, \prl {\bf
88}, 180401 (2002).

\bibitem{OLS98} The fractionilization could also be
realized with tightly confined 1D bosonic atoms in the Tonks gas
regime, where the impenetrable atoms obey FD statistics, M.
Olshanii, Phys. Rev. Lett. {\bf 81}, 938 (1998).

\bibitem{lattices} For recent experiments on bosonic atoms in optical lattices
see, e.g., C. Orzel {\it et al.}, Science {\bf 291}, 2386 (2001);
M. Greiner {\it et al.}, Nature {\bf 415}, 39 (2002).

\bibitem{BEL83} J.S. Bell and R. Rajaraman, Nucl. Phys. {\bf B220}, 1
(1983).

\bibitem{KIV} S. Kivelson and J. R. Schrieffer, Phys. Rev. B {\bf
25}, 6447 (1982); R. Jackiw {\it et al}, Nucl. Phys. B{\bf 225}, 233
(1983).

\bibitem{GW} J. Goldstone and F. Wilczek, Phys. Rev. Lett. {\bf 47}, 986
(1981).

\bibitem{JAV95} J. Javanainen and J. Ruostekoski, Phys. Rev. A {\bf 52},
3033 (1995).

\bibitem{AND96} M. R. Andrews {\it et al.}, Science {\bf 273}, 84 (1996).

\bibitem{ION80} D. J. Wineland, W. M. Itano and J. C. Bergquist, Opt.
Lett. {\bf 12}, 389 (1987).

\bibitem{DUM02} R. Dumke {\it et al.},
\prl {\bf 89}, 220402 (2002).

\bibitem{STA99} D.M. Stamper-Kurn {\it et al.},  Phys. Rev.
Lett. {\bf 83}, 2876 (1999); J. Steinhauer {\it et al.},  {\it
ibid.} {\bf 90}, 60404 (2003).

\end{references}
\end{document}